\documentclass{aa}
\usepackage{graphicx}
\usepackage{txfonts}
\def\({\left(}
\def\){\right)}
\def\[{\left[}
\def\]{\right]}
\begin{document}
\title{A cool disk in the Galactic Center?}
\author{B.F. Liu , F. Meyer  \and E. Meyer-Hofmeister}
\offprints{Bifang Liu}
\institute{Max-Planck-Institut f\"ur Astrophysik, Karl-
Schwarzschildstr.~1, D-85740 Garching, Germany}

\date{Received: / Accepted:}
\abstract
{We study the possibility of a cool disk existing in the Galactic
Center in the framework of the disk-corona evaporation/condensation model.
Assuming an inactive disk near the gravitational capture
distance left over from an earlier evolutionary stage,
a hot corona should form above the disk since there is a continuous
supply of hot gas from stellar winds of the close-by massive stars.
We study the interaction between the disk and the corona. Whether the
cool disk can survive depends on the mass exchange between disk and
corona which is determined by the energy and pressure balance.
If evaporation is the dominant process and the rate is larger than the
Bondi accretion rate in the Galactic Center, the disk will be depleted
within a certain time and no persistent disk will exist. On the other
hand, if the interaction results in hot gas steadily condensing into
the disk, an inactive cool disk with little gas accreting towards the
central black hole might survive in the Galactic Center. For this case
we further investigate the Bremsstrahlung radiation from the hot corona and
compare it with the observed X-ray luminosity. Our model
shows that, for standard viscosity in the corona ($\alpha=0.3$), the mass
evaporation rate is much higher than the Bondi accretion rate and the
coronal density is much larger than that inferred from Chandra observations. An
inactive disk can not survive such strong evaporation. For 
small viscosity ($\alpha \la 0.07$) we find condensation solutions.
But detailed coronal structure computations show that in this case
there is too much X-ray radiation from the corona to be in
agreement with the observations. From this modeling we conclude that
there should be no thin/inactive disk presently in the Galactic
Center.  However we do not exclude that the alternative non-radiative
model of Nayakshin (2004) might instead be realized in
nature and shortly discuss this question.

}

\maketitle
\section{Introduction}
The case for a massive black hole in our Galactic Center coincident
with the radio source Sgr A$^*$ represents a unique opportunity to probe the
dynamics of gas accreting onto a massive black hole. Dynamical
measurements of stellar velocities within the central 0.1 parsec of the
Galactic Center from 10 years of high resolution imaging indicate a
central mass of $2.6\times10^6M_\odot$ (Genzel et al. 2000, Ghez et
al. 2000) or
$3.3\times10^6M_\odot$ (Sch\"odel et al. 2003), recent
measurements of stellar orbits (Genzel et al. 2003) provide evidence
for a $3.6\times10^6M_\odot$ black hole.
The Eddington luminosity for
$3.3\times10^6M_\odot$ is $L_{\rm Edd}\approx 10^{44.6}{\rm erg\, s}^{-1}$.
Observations show that Sgr A$^*$ is an extremely dim galactic  nucleus.
The luminosity in the submillimeter/far infrared region is
$L_{\rm submm}\sim 10^{-8.6}L_{\rm Edd} $, even less in the infrared band
with $L_{\rm IR}< 10^{-9.6}L_{\rm Edd} $, also very low is the quiescent X-ray
luminosity, $L_{\rm X}< 10^{-11}L_{\rm Edd}$ (Narayan 2002,
where $L_{\rm Edd}=10^{44.6}{\rm erg\, s}^{-1}$).
Chandra observations directly image the hot X-ray-emitting thermal gas in
the vicinity of the Bondi accretion radius where the surrounding gas
is captured by the gravitational pull of the central black hole,
and determine temperatures and densities that allow to estimate a mass
accretion rate of Sgr A$^*$  of $\dot M_{\rm{Bondi}}\sim (0.3-1)\times
10^{-5}M_\odot$/yr, equivalent to$\sim 10^{-4}\dot M_{\rm
Edd}$ (with $\dot M_{\rm Edd}\equiv L_{\rm Edd}/0.1 c^2$)
(e.g. Baganoff et al. 2003).
If mass flows steadily through a Shakura-Sunyaev thin disk (Shakura \&
Sunyaev 1973) to the
central black hole at this accretion rate, the thin disk model would predict a
luminosity $L_{\rm disk}\sim 0.1 \dot M_{\rm Bondi} c^2\sim
10^{40.8}{\rm erg\, s}^{-1}\sim 10^{-4}L_{\rm Edd} $, much higher than
the luminosity in any band observed. Thus, a standard thin disk model
appears to be ruled out.

The low luminosity was explained by an advection-dominated accretion
flow (ADAF), with a spectral fit first presented by Narayan et al. (1995).
In the following years important observational results on the emission
of Sgr A* and theoretical work lead to an improved model for
radiatively inefficient accretion flows (RIAFs). RIAF models for Sgr
A* are discussed in the recent reviews of Yuan, Quataert \& Narayan (2003)
and Quataert (2003). According to these
models, most of the thermal energy released by viscosity and increased by
compression is retained in the gas and advected to the
central black hole. The RIAF model in addition assumes that very
little mass from large radii actually accretes onto the black hole
while a large part is lost through outflows during the accretion. These
models naturally yield the observed spectra of Sgr A$^*$. A key
constraint on these models is that the fraction
of gravitational energy heating the electrons must be very small and
hence a two-temperature treatment of the plasma is required.

Another possibility to explain the low luminosity of Sgr A* might come
from the existence of cold molecular gas in the parsec region
of the Galactic Center in the form of an inactive/dead thin disk
without accretion. Falcke \& Melia (1997) had suggested such a ``fossil''
disk (for a review see Melia \&
Falcke 2001). The model assumes that gas captured at Bondi accretion rate
condenses onto such an inactive disk without mass
accretion onto the black hole. This runs into
difficulties because the inflowing gas produces a fair amount of
luminosity in the infrared (Narayan 2002) as it clashes onto the disk
and loses its thermal and kinetic energy. This radiation is not seen.
Nayakshin (2004) revisits the concept of an inactive disk.
He suggests that, in the case of an extraordinarily long
mean free path (larger than the pressure scale of  the corona)
and extremely low  viscosity of the hot gas, the energy can
be conducted to a very thin transition layer by free streaming
electrons and is then radiated in infrared to UV wavelengths. As this
radiation from the thin layer would be observed edge-on a discrepancy
between predicted and observed luminosities could be avoided.
This interesting suggestion deserves further analysis of the
transition between the hot gas and the cool layer and the
coupling between the hot ions and the energy transferring electrons.

As a further contribution to the issue of a cool disk around the
Galactic Center we here study the vertical structure of a hot corona
in interaction with a disk below. We assume a cool disk in the outer
region around the circularization radius where the free fall of a Bondi type
accreting hot gas goes over into a circular motion around the
gravitational center due to its specific angular momentum.
Such a disk might not be unreasonable since the system must have been
quite bright during an earlier evolutionary stage and angular momentum that was
released by the high accretion rate should have moved disk mass into
outward regions. With winds from young stars being captured by
the gravitational field of the black hole, a corona unavoidably forms above
the cool disk. The question is, how do disk and corona evolve?
Does the hot gas condense to the disk with little mass actually
accreting to the black hole, or does mass rather evaporate from the disk
to the corona overwhelming the incoming hot gas and even finally
depleting the cool disk underneath?

The answer depends mainly on the
rate at which the hot gas is supplied from the capture radius.
If gas is supplied to the corona at a sufficiently high rate coronal
gas condenses to the cool disk. If no gas is supplied from the outside
or if the outside gas supply is too small mass instead evaporates from the
disk into the corona. Both processes are the
consequence of pressure and energy equilibrium between the disk and
the corona (Meyer et al. 2000, Liu et al. 2002). Here we
study in detail the structure of the corona for the case of the Galactic
Center in order to see what the dominant process between disk
and corona is, condensation or evaporation.  Can the cool disk survive if
evaporation is dominant?  Furthermore, the detailed computation
allows to calculate the Bremsstrahlung luminosity of the corona and compare it
with the observed X-ray luminosity. We show that these
results exclude the existence of any cool disk in our Galactic Center.

In Sect.2 we describe the physics of the interaction between the disk and
the corona. In particular we discuss how a radial inflow of mass from the
outside affects the mass and energy balance in such a corona.
In Sect.3 we present numerical results and show how the
value of the viscosity affects evaporation or condensation.
We discuss several aspects of our results in Sect. 4, including 
 a comparison with a non-radiative condensation model 
of Nayakshin (2004) and the question of a past disk 
being evaporated now in the
Galactic Center. A conclusion follows in Sect. 5.

\section{The physics of interaction between disk and corona}

For a hot corona lying above a cool disk, interaction between
the disk and the corona occurs via  energy and mass exchange.
The hot corona conducts heat downward by electrons. At the
bottom the heat is radiated away. If the density in the corona is too
low, Bremsstrahlung is inefficient and the thermal conductive flux
heats up some of the disk gas leading to mass evaporation from the
disk into the corona. The resulting density increase in the corona
raises the radiation loss and thereby counteracts further
evaporation. If the coronal density is too high, radiative cooling is
too strong and gas condenses into the disk. At the final
equilibrium density, cold gas steadily evaporates from the disk into
the corona if mass is drained continuously from the corona inward by
diffuse flow, or hot gas steadily condenses to the disk
if the corona continuously gains mass by mass flow. For
example, when there is no hot gas coming in through the outer
boundary, (case 1), mass is continuously lost from the
corona by accretion towards the central object. This is resupplied by
evaporation from the surface of the cool disk as the corona tries to
restore the density to the equilibrium level. If there is more hot gas
being fed in at the outer boundary than what flows inward towards the
center, (case 2), hot gas continuously condenses to the cool disk.

\subsection{Basic equations}
The equations describing the corona above the disk are derived in
our earlier studies (Meyer \& Meyer-Hofmeister 1994; Meyer et al. 2000;
Liu et al. 2002). For the Galactic Center, we consider the region around the
circularization  radius $10^4-10^5 R_{\rm S}$ ($R_{\rm S}$
Schwarzschild radius) far from the black hole.
In this region, electrons are collisionally
well coupled to ions, temperatures of electrons and ions are the same.

We now discuss the upper boundary condition for the corona at such
radii. The earlier investigations showed that in general wind loss
from the corona is an integral part of the solution, where a sonic
transition occurs at some height and the wind flow cross section
flares out with the wind expansion. In an advanced multi-zone
modeling (Meyer-Hofmeister \& Meyer 2003) we found that the wind pressure
is highest at the distance where the evaporation efficiency is highest,
at a few hundred Schwarzschild radii. The pressure in the expanding
wind from this region even dominates over the pressure at a sonic
point of winds from farther out regions and prevents sonic
transition and wind loss there altogether. Thus for evaporation
solutions in these outer regions we apply the condition of zero
vertical velocity at a height of $z=R$. The actual height at this
point is not important as long as it includes the lower down region
where most of the coronal action occurs.

In the case of condensation
solutions the incoming mass flow from the outside accretion anyhow
dominates the coronal pressure and prevents any free wind
expansion. We thus can apply the same boundary condition. In our
equations we  can also leave out the so called flaring terms which are
only important in the wind expansion geometry.

In the following we
list the four  ordinary differential equations describing the
coronal flows above a disk in the Galactic Center.

Continuity of mass flow
\begin{equation}\label{e:cont}
{d\over dz}(\rho v_z)={2\over R}\eta_M \rho v_R.
\end{equation}
$z$-component of the equation of motion
\begin{equation}\label{e:mdot}
\rho v_z {dv_z\over dz}=-{dP\over dz}-\rho {GMz\over (R^2+z^2)^{3/2}},
\end{equation}

Energy equation
\begin{equation}\label{e:energyt}
\begin{array}{l}
{d\over dz}\[\rho v_z\({v^2\over 2}+{\gamma\over\gamma-1}{P\over\rho}
-{GM\over\(R^2+z^2\)^{1/2}}\)+F_c\]\\
={3\over 2}\alpha P\Omega-n_e n_iL(T)\\
+{2\over R}\eta_E \rho v_R\({v^2\over 2}+{\gamma\over\gamma-1}{P\over\rho}
-{GM\over\(R^2+z^2\)^{1/2}}\).
\end{array}
\end{equation}

Thermal conduction for a fully ionized plasma
\begin{equation}\label{e:fc}
F_c=-\kappa_0T^{5/2}{dT\over dz}
\end{equation}
with $\kappa_0=10^{-6}$ ergs ${\rm s}^{-1} {\rm cm}^{-1}{\rm K}^{-7/2}$ (Spitzer 1962).

Here $\rho, P, T$ are density, pressure and temperature, $v_R$ and $v_z$
the radial and vertical velocity, $M$ is the black hole mass, $G$
the gravitational constant and $\Omega$ the rotational frequency,
$n_e$ and $n_i$ are electron and ion particle densities, $n_e n_iL(T)$ is the
Bremsstrahlung cooling rate, and $\gamma$ the ratio of specific heats.
$\alpha$ is the viscosity parameter (ratio of viscous stress to
pressure).
The terms $\eta_M$ and $\eta_E$ account for radial mass and energy
flows with $\eta_E=\eta_M+0.5$, explained in the next section.
 The difference between $\eta_M$ and
$\eta_E$ results from the fact that
the specific energy which the mass flow carries scales radially as
$\frac{1}{r}$.
We consider stationary azimuthally symmetric flows.

Compton cooling is negligible for the coronal 
structure at distances of $10^4$ to $10^5$ Schwarzschild radii.
Three possible contributions to Compton cooling have to be considered. 
(1) The cooling by the radiation from the disk surface caused by mass flow 
in the disk was investigated by Liu et al. (2002) and shown to be negligible
even for mass flow rates in the disk of $0.02 \dot M_{\rm {Edd}}$. 
(2) We have estimated that the energy loss by Compton cooling from 
radiation caused by reprocessing of coronal X-rays is always less than 
$0.3\%$  of that by bremsstrahlung. (3) The Compton effect from the X-rays 
from the central source and the surrounding gas is negligible 
in our context because of the very low observed radiation.

At the lower boundary $z_0$ we start our calculations at the
temperature $T=10^{6.5}$K. 
This value is in the steep temperature profile in a thin transition 
zone of nearly constant pressure. Its physics can be described by the 
balance between gain of heat by thermal conduction and radiation loss.
Other effects like frictional heating and energy transport by the vertical 
mass flow are negligible here (Meyer et al. 2000). This establishes a 
relation between temperature and heat flux which can be scaled according 
to the pressure (Smeleva \& Syrovatskii 1973). If one temperature in the 
profile is selected then, due to the scaling, one obtains a unique 
relation between thermal heat flux and pressure (Liu et al. 1995).
We use this relation
\begin{equation}
 F_c=-2.73\times 10^6 P \rm{\ in\ cgs\ units}.
\end{equation}
As discussed above at the upper boundary $z=R$ we take
\begin{equation}
F_c=0\ {\rm and}\   v_z=0.
\end{equation}

\subsection{Parameter $\eta_M$ for the net inflow of mass}
The parameters $\eta_M$ and $\eta_E$ are  introduced to account for the radial mass
and energy flows (Meyer-Hofmeister \& Meyer 2003). In the early
evaporation model (``one-zone-model'', Meyer \& Meyer-Hofmeister 1994), it is
assumed that essentially all the mass evaporating from the disk to the corona
flows towards the farther inward coronal region and that there is no
significant mass flow into the corona from outside.
In our designation this is $\eta_M=1$. Taking into account the
structure of the radially neighboring
coronal areas and the mass and energy exchange with these we have
to consider the actual radial gradients of mass flow in the vertical
structure equations. The introduction of $\eta_M$ is  to parameterize
the unknown radial gradient in the  general mass conservation
equation,  ${\partial\over \partial z}(\rho v_z)+{1\over R}{\partial
 \over \partial R}(R\rho v_R)=0$. We approximate this in
Eq.(\ref{e:cont}) with a height-averaged factor $\eta_M$, which implies,
\begin{equation}\label{e:eta1}
\eta_M\approx {\dot M(R_{in})-\dot M(R_{out})\over \dot M(R)},
\end{equation}
where $\dot M(R_{out})$ is the incoming mass flow rate from the
outer boundary of the one-zone corona, $\dot M(R_{in})$
mass outflow rate at the inner boundary, and  $\dot M(R)$ the
typical mass flow rate in the one-zone corona. Thus, the parameter
$\eta_M$, by its definition, depends on the net mass gain/loss
rate through the radial boundaries.

Integration of Eq.(\ref{e:cont}) along the $z$-direction gives,
\begin{equation}\label{e:eta2}
\eta_M \dot M(R)=2\pi R^2\dot m_0,
\end{equation}
where $\dot m_0$ is mass evaporation rate ($\dot m_0>0$) or
condensation rate ($\dot m_0<0$)
per unit area at the interface between disk and corona.
Obviously, $\eta_M>0 $ describes the fraction of the coronal mass flow
contributed by disk evaporation, $\eta_M<0$ describes the fraction of
the coronal flow condensing to the disk. A special case $\eta_M=0$
means there is no mass evaporation or condensation between disk
and corona. If there is any gas coming from the outer boundary, all of
this flows through the corona towards the central black hole.
Two extreme cases are $\eta_M=1$  and $\eta_M=-1$. The former
represents the case when no mass enters through the outer boundary and
all the mass flowing in the corona is contributed by the
evaporation, $\dot M(R)=2\pi R^2\dot m_0=\dot M(R_{in})$.
The latter case represents the situation in which no mass flows towards
the central object at the inner boundary and all the mass entering the corona
from the outer boundary condenses to the disk,
$\dot M(R)= \dot M(R_{out})=-2\pi R^2\dot m_0$.
Therefore, our parameter $\eta_M$ lies in the range
$-1\le\eta_M\le 1$.

\section{Condensation or evaporation of gas in the central arcsecond
of our Galactic Center?}

In our Galactic Center, hot gas in the
central parsec produced by winds from massive stars is
gravitationally captured by the black hole at a distance around 0.04
parsec which corresponds to 1 arcsecond at the sky. If
there is a cool (inactive) disk in this region, a corona forms above
the disk, fed by the captured hot gas. The situation is a bit
different from the one discussed in our previous evaporation model,
where we assumed
that cold gas flows inward from the outer region (e.g. a secondary star)
via the thin disk and then feeds the corona through evaporation. There
it was assumed that essentially no significant hot gas comes from a
further outward located hot corona.

Now the situation is turned around, hot
gas accretes through the corona without any cold gas being fed
into the cool disk from the outer disk boundary.
As explained in the last section, if much mass is fed into the corona,
the pressure in
the corona is high. The pressure and energy balance between
disk and corona leads to mass condensing from the corona to the
disk. On the other hand  if the feeding rate to the corona is low, mass
evaporation from the disk to the corona is the dominant process.
Strictly speaking, how large the mass exchange rate is and
whether mass evaporates or condenses depend on the rate of
net mass inflow into the coronal region, i.e., the difference between the
mass feeding rate at the outer boundary and the mass loss rate to
the central object at the inner boundary.

If the equilibrium between
disk and corona leads to strong mass evaporation, the cool disk will
eventually be depleted and mass then accretes via a hot corona/ADAF or
RIAF. If condensation dominates, the hot gas captured by the
gravitational field of the black hole finally flows to the disk and
is deposited there with little mass actually accreting onto the black hole.
In the following we investigate these coronal features for the case of
our Galactic Center.

\subsection{Standard viscosity $\alpha=0.3$}
At the Galactic Center, hot gas accretes at a Bondi accretion rate of
$\sim 10^{-4}\dot M_{\rm Edd}$ from the capture radius
$R \sim 10^5 R_{\rm S}$ (0.8 arcseconds),  
with the circularization radius  estimated
as $R \sim 10^4 R_{\rm S}$. The circularization radius might also be larger, 
e.g. if the accreted winds predominantly come from close-by stars 
orbiting in a common sense around the Galactic Center.
We thus investigate the coronal
features  in a  region from $10^4 R_{\rm S}$ to  $10^5 R_{\rm S}$
according to the scenario described in the section above.
For our computations we took the mass of the black hole as $3.3\times
10^6M_\odot$.
We first discuss the results for $\alpha=0.3$. 
For this value our earlier modeling of hard/soft spectral transitions
in X-ray binaries and applications to AGN gave good quantitative
comparisons with observations. Henceforth we call it the "standard
value". A similar
value was also chosen for the application of ADAFs to luminous black
hole X-ray binaries (Esin et al. 1997).

\subsubsection{Evaporation model}

\begin{figure}
\centering
\includegraphics[width=7.3cm]{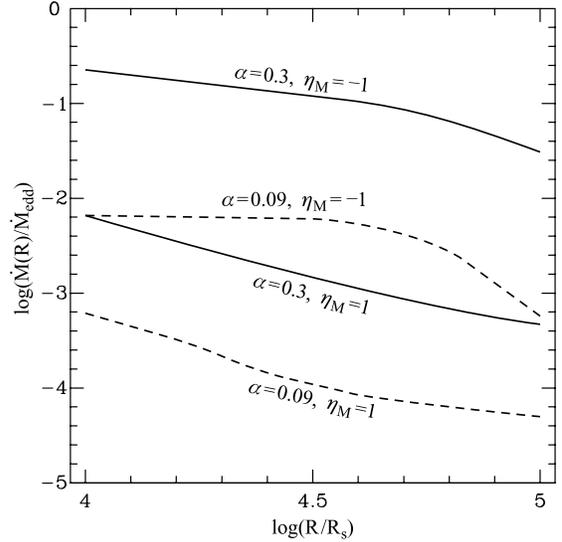}
\caption[]{\label{f:dm-r} Radial mass flow rate in the corona for
given parameters  $\eta_M=1$, evaporation, and $\eta_M=-1$,
condensation; Solid lines:
solutions for standard viscosity, $\alpha=0.3$, dashed lines
solutions for a lower viscosity $\alpha=0.09$.
Note that the curves do not represent a distribution of the radial
mass flow in the corona, but give the typical mass flow rate
in the local corona at the distance $R/R_{\rm S}$ under the
conditions specified with $\eta_M$.}
\end{figure}

In our terminology any value of $\eta_M>0$ means there is more gas
flowing out of the coronal region than coming in. The net mass flow
out of the corona is contributed by mass evaporation from the disk.
The case of ``standard evaporation'', considered in the
one-zone-model  corresponds to $\eta_M=1$, which means that the
typical mass flow in the corona is the same as the mass
evaporation into the corona at that distance.

Fig.\ref{f:dm-r} shows the radial mass flow rate in the corona
(on both sides of the disk) for distances $10^4 R_{\rm S}$ to $10^5 R_{\rm
S}$. The mass flow rate is $\dot M(R)=-2{\int}_{z_0}^{z_1} 2\pi R \rho
v_R dz$,
with $v_R\approx -\alpha V_s^2/\Omega R$ the typical radial drift velocity
($V_s$ isothermal sound
speed, $\Omega$ Kepler angular frequency, $z_0$ and $z_1$
lower and upper boundary of the corona respectively).

The lower solid line represents the mass accretion rate
for  the standard one-zone evaporation model.
We see that the mass flow rates in the coronal region from
$10^4 R_{\rm S}$ to  $10^5 R_{\rm S}$
are around $10^{-3}\dot M_{\rm Edd}$,  about 10 times higher than the
incoming mass flow estimated from Bondi accretion and Chandra
observations as $\dot M\sim 10^{-4}\dot M_{\rm Edd}$.
This indicates that the dominant process in the disk-corona system is that
gas evaporates from the disk to the corona and then flows towards the
central black hole. The hot gas captured at the Bondi radius is only
a minor contribution and hardly affects the coronal structure.
A strong corona with high density and high temperature
is built up above the disk by the mass evaporation.

\begin{figure*}
\centering
\includegraphics[width=14cm]{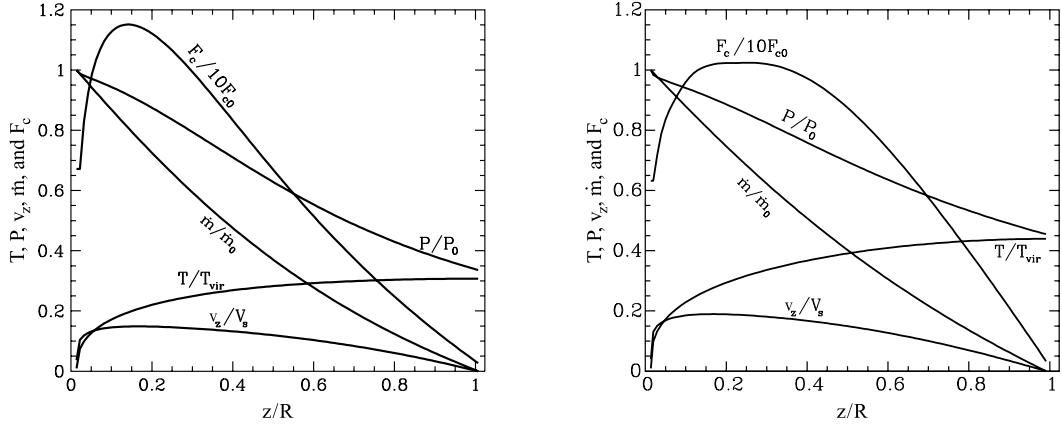}
\caption[]{\label{f:vert} Vertical structure of the corona with mass
evaporation from the disk to the corona  ($\eta_M=1$, left panel) and
with mass
condensation from the corona to the disk ($\eta_M=-1$, right panel)
at the distance $R=10^4 R_{\rm S}$ from a central black hole of
$3.3\times 10^6M_\odot$. $z/R$ ratio of height above midplane to distance.
Distribution of the following quantities: Temperature $T$ in units
of virial temperature $T_{\rm {vir}}$, pressure $P$,
thermal conductive flux $F_c$ and vertical mass flow rate $\dot m=\rho v_z$
(the latter ones scaled to their values at the lower boundary:
$P_0\approx 1.9\times 10^{-2}{\rm dyn}\,{\rm cm}^{-2}$,
$F_{c0}\approx -5.2\times 10^4$ ergs ${\rm cm}^{-2}{\rm s}^{-1}$ ,
$\dot m_0\approx 3.4\times 10^{-11} g\,{\rm cm}^{-2}\,s^{-1}$  for the left panel, and
$P_0\approx 0.6\,{\rm dyn}\,{\rm cm}^{-2}$,
$F_{c0}\approx -1.6\times 10^{5}$ ergs ${\rm cm}^{-2}{\rm s}^{-1}$,
$\dot m_0\approx -10^{-9} g\,{\rm cm}^{-2}\,s^{-1}$ for the right panel
).}
\end{figure*}

Fig.\ref{f:vert}
shows the coronal structure at $R=10^4 R_{\rm S}$
for both cases, evaporation and condensation. From the figure
we see that  in the case of evaporation the typical temperature of the
coronal gas is $\sim 0.3 T_{vir}\sim 10^8$K (virial temperature
$T_{vir}=GM/(R\frac{\Re}{\mu})$, $\Re$ gas constant, $\mu$ molecular
weight, taken as 0.62). The particle number density which follows
from pressure and temperature is larger than
$5\times 10^5{\rm cm}^{-3}$. Temperature and density both are much larger than
the values observed by Chandra at the Galactic Center.
A disk might have existed earlier but it could have been
eventually depleted by the mass evaporation since there is no mass
supply for the cool disk. Therefore, the
standard disk-corona evaporation model appears to exclude the continued
existence of a cool disk in the Galactic Center. Note that this
argument is independent on whether the cool disk is completely inert or
whether it allows some mass flow itself.

\subsubsection{Condensation model}
To let coronal gas condense to the disk, we need high pressure in
the corona. This requires  more gas feeding into the corona from outside
than gas flowing out of the corona at the inner boundary.
The value of $\eta_M$ is then negative. An extreme case we are
interested in for our investigation is that all the gas coming in from
the outer boundary condenses to the disk without any coronal accretion
toward the central object, $\eta_M=-1$.
The mass flow rates  in the coronal region at distances $10^4 R_{\rm
S}$ to $10^5 R_{\rm S}$ are shown in  Fig.\ref{f:dm-r}
as upper solid line. Obviously, the derived mass flow rate
$\dot M(R_{out})=-\eta_M\dot M(R)=\dot M(R)$ with condensation
($\eta_M=-1$ prescribed), is higher than that with mass evaporation
($\eta_M=1$ prescribed).

The condensation rate is far too high
compared to the Bondi accretion rate in the Galactic Center.
The density in such a corona is also much higher than the value
inferred from the observations. Therefore, we cannot expect that, if there is
a cool disk in the Galactic Center, hot gas captured by the black hole
at the Bondi radius mainly condenses into the disk and is deposited
there. Instead, the disk gas will evaporate into the
corona, increasing the coronal flow inward significantly.
As noted above, this process can finally deplete the cool disk.
From then on no cool disk exists anymore
in the Galactic Center.

\subsubsection{Dependence of the solutions on $\eta_M$}

\begin{figure}
\centering
\includegraphics[width=7.3cm,clip]{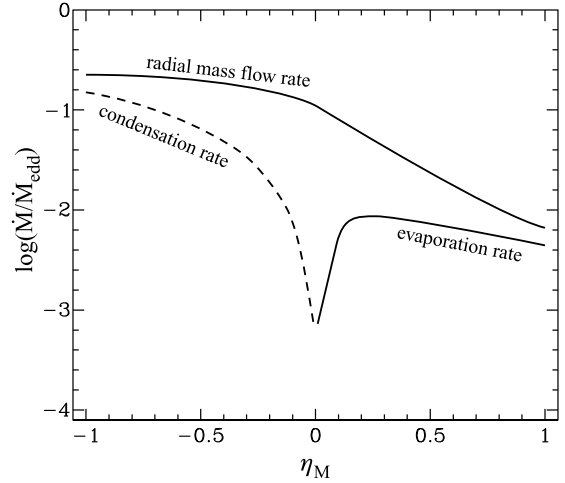}
\caption[]{\label{f:dm-eta} Radial
mass flow rate in the corona (upper curve) together with
condensation rate$=-2\pi R^2 \dot m_0/\dot M_{\rm Edd}$ (dashed line), and
evaporation rate$=2\pi R^2 \dot m_0/\dot M_{\rm Edd}$ (lower solid line), for the
Galactic Center at $R=10^4 R_{\rm S}$ (viscosity parameter
$\alpha=0.3$).
At $\eta_M$=-1 all mass that flows in the corona has
come from the outside and condenses into the cool disk, then with
increasing  $\eta_M$ a smaller part settles into the disk until at
$\eta_M$=0 no mass condenses. At $\eta_M$=1 all mass flowing inward in
the corona has evaporated from the cool disk (for details see text).
}
\end{figure}

More generally, we study the condensation and
evaporation solutions for a sequence of values of  $\eta_M$.
Fig.\ref{f:dm-eta} shows how the mass flow rate and the
evaporation/condensation rate at $R=10^4R_{\rm S}$ change with the
value of the parameter $\eta_M$. The results show that condensation
solutions are connected with a high mass flow rate through the
coronal region. At $\eta_M=-1$ all mass that flows in the corona has
come from the outside and condenses into the cool disk. With increasing
$\eta_M$ a smaller and smaller part of the coronal mass flow settles
into the disk until at $\eta_M$=0 no mass condenses and all mass that
comes from the outside continues inward. For increasing positive
$\eta_M$ mass evaporation contributes a growing fraction to the mass flow in
the corona until at $\eta_M$=1 all mass flowing inward in the corona
comes from evaporation of the cool disk. (The difference between upper
and lower curve at $\eta_M$=-1 and 1 is
an artifact of the one-zone model approximation and results from the
different way in which averages over the one-zone area are defined.)
One may note that as the characteristic mass flow rate is proportional to the
pressure in the corona, $\dot M(R)=-2\int_{z_0}^{z_1} 2\pi R \rho
v_R dz\propto \int_{z_0}^{z_1}Pdz$, the results in Fig.\ref{f:dm-eta}
reflect the intrinsic relation: More pressure means more radiative
cooling which supports condensation, less pressure means less
radiative cooling and supports evaporation.

Thus, if we know the pressure (or mass content) in a coronal region,
our computations for the equilibrium between disk and corona  allow
to derive how much mass condenses or evaporates. This is not directly
determined by the incoming mass flow alone, but depends on the net mass flow
into the region, i.e.,
the difference of incoming mass flow at the outer boundary and the
outgoing mass flow at the inner boundary. Therefore, when we
know  both the inner and outer boundary conditions as in multi-zone
modeling (Meyer-Hofmeister \& Meyer 2003), we
are able to determine the vertical and the radial structure of the
corona. Otherwise, $\eta_M$ will be an open parameter in the range
of -1 and 1.

Combining Fig.\ref{f:dm-r} and Fig.\ref{f:dm-eta}, we find that, for
$-1\le\eta_M\le1$, the mass flow rate in the corona is between the
two solid curves of Fig.\ref{f:dm-r}. The same is true for all
distances of $10^4 R_{\rm S}\la R\la 10^5 R_{\rm S}$.  In other words,
no matter how large a fraction of the coronal mass flow
condenses to the disk ($\eta_M$ between -1 and 0),
or how large a fraction of the coronal mass flow originates from
disk evaporation ($\eta_M$ between 0 and 1),
the mass flow in the corona would be much larger
than the Bondi accretion rate of $10^{-4}M_{\rm Edd}$ inferred from
observations. The detailed calculations also show that the hot gas
density is much larger than that obtained from the observations.

As a consequence of such a large mass flow rate, the luminosity of
coronal radiation in X-rays (note that it is not in the infrared),
estimated from
$L \sim GM\dot M(R)/2R=2.5\(R/R_{\rm S}\)^{-1}\[\dot
M(R)/\dot M_{\rm Edd}\]L_{\rm Edd}$,
is very high. For instance, at
$R\sim 10^5 R_{\rm S}$, $\dot M(R)\sim 10^{-3}\dot M_{\rm Edd}$, the
radiation from the corona would then be $L\sim 2.5\times 10^{-8}L_{\rm Edd}$,
much larger than the observed quiescent X-ray luminosity $L_{\rm
X}\sim 10^{-11} L_{\rm Edd}$. The Bremsstrahlung radiation is
confirmed by detailed computations of the coronal structure.

Therefore, for standard $\alpha$, these consistent disk-corona model
calculations exclude a disk at distances
$10^4 R_{\rm S}\la R\la 10^5 R_{\rm S}$ from the Galactic Center by
comparison of mass flow rate, density, and luminosity of the hot
gas predicted by theory with those actually observed.

\subsection{Low viscosity}

In the last section we showed that for standard $\alpha$ no
condensation solution exists that is compatible with the observed
mass accretion rate in the Galactic Center.
We investigate whether agreement between
observation and a disk-corona analysis can be achieved for smaller
values of $\alpha$.

In the models the viscosity parameter enters directly into the radial
drift velocity and the viscous release of heat, both of which become
smaller with smaller $\alpha$. Since radiative cooling is not affected
by this change in  $\alpha$ the balance between heating and cooling
requires less pressure in the corona. As a consequence one expects
that pressure, temperature, and radial mass flow in the corona
decrease when $\alpha$ becomes smaller (Meyer-Hofmeister \& Meyer 2001).

This effect is shown in Fig.\ref{f:dm-r} where the dashed lines give
the mass flow rates in the corona for a smaller viscosity
$\alpha=0.09$.
The upper and lower dashed lines show the condensation
($\eta_M=-1$) and the evaporation solutions ($\eta_M=1$) for
$10^4\le R/R_{\rm S} \le 10^5$.
For other values of $\eta_M$,  $-1<\eta_M< 1$,
the mass flow rates lie between these two lines. Indeed, a small
viscosity in the corona results in a significantly decreased mass flow
rate. Density and pressure in the corona also are smaller.
For very small $\alpha$, the mass flow rate in the corona can then, in
principle, become comparable with the capture rate
of $\sim 10^{-4}\dot M_{\rm Edd}$.

\subsubsection{Evaporation model}
In the case of evaporation, $\eta_M>0$, the mass flow rate  $\dot M(R_{in})$
leaving the coronal region at its inner boundary is always larger than
the incoming flow rate $\dot M(R_{out})$ at its outer boundary. Thus
an energy of the order of 1/2 the gravitational potential
energy of the accretion rate is released. Since the
temperature of the corona is some tens of a percent of the virial temperature,
i.e. $T\sim 10^8$K at the distance $R\sim 10^4 R_{\rm S}$ and
$T\sim 10^7$K at the distance $R\sim 10^5 R_{\rm S}$, the
coronal radiation is in the X-ray range and the corona should have a
luminosity $L_X\sim 2.5\times 10^{-9} L_{\rm
Edd}$ as long as there is hot gas flowing inward at the Bondi accretion rate
$\dot M=10^{-4}\dot M_{\rm Edd}$ at $R=10^5 R_{\rm S}$ and higher at
smaller radii. This is 2
orders of magnitude larger than the observed X-ray luminosity. Thus,
for any small $\alpha$, the disk evaporation solution can not be
consistent with the observations.

\subsubsection{Condensation model}
For condensation solutions part or in the extreme case all of the
incoming mass flow $\dot M(R_{out})$ would settle through the corona
into the cool disk and only a fraction or nothing at all would flow
inward through the inner boundary. In any of these cases again an
amount of energy of the order of 1/2 gravitational
potential energy of the incoming gas is released as heat. One thus
has a similar large discrepancy between observed and predicted
luminosities.

If $\dot M(R_{out})$ is the Bondi accretion rate in the
Galactic Center, $\dot M(R_{out})=10^{-4}\dot M_{\rm Edd}$,
consistency with a condensation solution ($-1<\eta_M< 0$) can
only be obtained by assuming a  small viscosity parameter
$\alpha$ (for example, $\alpha=0.07$ for $\eta_M=-1$ at
$R=10^{5}R_{\rm S}$ and an even smaller value at smaller distances).
However, in either of the extreme cases, $\dot M(R_{in})\approx \dot
M(R_{out})$ or $\dot M(R)\approx \dot M(R_{out})$, or any case between
those, $-1< \eta_M< 0$, our detailed computations of the coronal
structure show that the X-ray luminosity produced by Bremsstrahlung
radiation in the corona by far exceeds the luminosity observed. The
existence of a cool disk in the Galactic Center seems thus to be
excluded. For a condensation picture this was already pointed out
by Narayan (2002).

In addition to this result for the situation in the Galactic Center
with the presently observed mass accretion rate the results of our 
investigation allow to discuss whether a formerly existing 
accretion disk in the context of gravitational instability and star 
formation in this disk has disappeared due to evaporation until now.

\section{Discussion}
\subsection{Coronal winds}
We have here investigated solutions for which no wind escapes from the corona. 
This is e.g. the case if winds from farther-in coronal regions fill 
the "coronal sky" with pressure that prevents the escape of winds from 
outer regions. We have estimated that this is the case already for winds that 
originate from the innermost cool disk region at $R=2\times 10^3$ 
Schwarzschild radii (the distance at which a presumptive disk heated only
by its coronal energy release is still cool enough to prevent a magnetic 
dynamo and magnetic friction).

Should a presumptive cool disk however be cut off already at about $10^4$ 
Schwarzschild radii the coronal solution has to allow for the escape 
of a wind (if not suppressed by the accreting gas ram pressure). 
Such a solution has already been obtained earlier (Meyer et al. 2000, for 
the case $\eta_M=+1$, "the weakest" coronal radiation case). In this case 
the corona in the upper layers is more tenuous and and the predicted X-ray 
radiation is somewhat less but with $L_{\rm X} \approx 10^{37.1}$ erg/s 
still far above the value of $10^{33.6}$ erg/s allowed by Chandra 
observations (Baganoff et al. 2003).

\subsection{Coupling of electrons and ions in the corona}
As shown in Sect.3 (see also Fig. \ref{f:dm-r})
a  small viscosity is
required if the density in the corona should not be much higher than
inferred from the observations. Our condensation solution shows
that, in this case (say, $\alpha=0.09$),
the temperature in the corona is about 1/3
 of the virial temperature and the gas density is fairly low.
In such a tenuous hot gas the mean free path can become quite
large. Thus the question arises whether the Spitzer formula
($F_c=-\kappa_0 T^{5/2}dT/dz$) can still be applied for the
thermal conduction and whether the electrons are well-coupled to the
ions so that temperature equilibrium between the two species is
established. In our vertical structure computations  we therefore
compare, at every step of the vertical integration, the
conductive heat flux $\kappa_0 T^{5/2}dT/dz$ with the thermal flux of
``free streaming'' electrons ${3\over 2}PV_s$.

We find that for our solutions, the thermal flux
given by the Spitzer formula never exceeds the saturated value for free
streaming electrons and the mean free path remains always less than
the thermal scale height ($T/(dT/dz$)). Since $F_c=0$ is required
at the upper boundary,
 as the density becomes smaller and
smaller in the upper layers, the temperature gradient approaches zero.
This results in a very small heat flux (from the Spitzer formula), which
is smaller than the ``free-streaming'' heat flux. Though the mean free
path is quite large at these upper layers it also remains smaller than
the temperature scale height which approaches infinity.

We also compared
the time needed for equipartition temperature between electrons and
ions with the thermal timescale and confirmed that
the former is always shorter than the latter. This means that
in such a corona, far from the central black hole, electrons and ions
are well coupled. Therefore, these models for the corona above a cool
disk around the Galactic Center are self-consistent.

\subsection{The effect of magnetic fields}
Could magnetic fields affect our solutions? The standard $\alpha$ 
description is generally taken as describing the effect of dynamo created 
magnetic fields, here in our case in the corona. Could additional fields 
from an underlying disk affect these solutions? Such an effect was discussed 
by Meyer and Meyer-Hofmeister (2002) for low luminosity AGN where it was 
shown that it can have a significant effect on the truncation of accretion 
disks. The temperature of the disk around Sgr A* from reprocessed coronal 
radiation, $<10^{2.8}$K, however, is too low to allow a dynamo to operate.

\subsection{Alternative model}

Recently, Nayakshin (2004) discussed an alternative model
which suggests that all the hot gas captured by the black hole is
deposited into an inactive disk (see also Nayakshin 2003).
In that model, the corona is extremely tenuous so that the electron
mean free path becomes large compared to the pressure scale height of
the corona. Thus, a ``free-streaming electron'' thermal conduction
law is applied instead of the classic Spitzer formula. The corona is
approximated as a homogeneous column on top of a very thin
transition region to the cool disk, in contrast to the vertically layered
corona discussed in this paper.
The free-streaming particles of long mean free path are thought to
transport the thermal energy into the cool disk at the saturation speed of
the order of the sound velocity in the corona, thus the coronal gas
would be able to sink down into the cool disk at a fraction of that
speed and condense as tenuous gas without significant radiation in the 
X-ray band, 
low enough not to get into conflict with the observed low X-ray luminosity.

This model lies in a quite different parameter space as the the one
discussed in our paper. It differs in the assumption of a very thin
transition between the tenuous hot corona and the dense cool disk
from the extended transition layer obtained in our model.

Assuming that both these different models are self-consistent the
question arises which one might be realized in nature. We have seen
that the addition of the hot tenuous accreting gas to the already
existing layered corona discussed in this paper only constitutes a
minor contribution to the dominant coronal evaporation process with
its significant radiation in the X-ray band. Such
coronal evaporation models were successfully used to explain the
formation of inner holes in quiescent accretion disks and the soft/hard
transition in spectra of soft transients and high-mass X-ray
binaries. On the other hand if the tenuous non-radiative accretion
would be set up before the standard evaporation could have established
itself it might have prevailed until now. This question requires
further investigation.

\subsection{A disk in the Galactic Center in the past?}
Assuming the applicability of our modeling we can estimate an upper
limit for the mass that might have been left over in a putative
accretion disk after a last star forming event, but
would now have evaporated by coronal action. No significant
X-ray luminosity would then have to be expected today. Such a scenario is
not unreasonable since accretion of mass to the central black hole
releases angular momentum that is transferred outside in form of matter
with Kepler specific angular momentum.

The evaporation rate depends on the value of the viscosity assumed.
For a standard value of $\alpha$=0.3 and dominant evaporation,
$\eta_M$=1, one has $\dot M/ \dot M_{\rm Edd}\sim 10^{-2.9}$
corresponding to $ \dot M\approx 10^{-4} M_\odot$/yr. Observations of
stars at the Galactic Center and especially spectroscopy of
one such star, S0-2, suggests that these are main sequence O/B stars
(Eisenhauer et al. 2003, Ghez et al. 2003).
The O/B stars close to the Galactic Center could not have formed
longer before their main sequence lifetime, of order of $10^{6.5}-10^7$yr
s, (Maeder \& Meynet 1989). Thus a disk that remained after the stars had
been formed but has evaporated by now should have contained gas not more
than 300 to 1000 $M_\odot$.
This value is in the range of mass of the presently
observed bright O/B stars close to the Galactic Center. Interestingly,
but, perhaps by accident, it is also close to the stability limit of a
disk against self-gravitation. For an evaporating disk
with effective temperature of 50K
at distance $10^{16.9}$ cm, the mass of the disk of the stability
limit (see e.g. Gammie 2001) is about $10^{3.4} M_\odot$.

Thus a disk that had become unstable by self-gravitation and formed
the presently observed young massive stars around the Galactic Center until
the gravitational instability had ceased could perhaps have now
completely disappeared by the process of coronal evaporation. But
this remains a rather speculative question until we know
more about the stellar population and its origin so close to the
Galactic Center.

\section{Conclusions}

We have developed a model for a cool disk around the Galactic
Center that has a corona above it, allowing for a coronal accretion of
gas captured at the Bondi radius from stellar winds of massive stars.
Our models incorporate hydrostatic
layering, thermal heat conduction, friction, and radiative energy
loss and allow for mass exchange between disk and corona by
condensation or evaporation of gas, as well as
mass gain or loss by radial flow in the corona (the latter a
generalization of the former one-zone model).

We find that the solutions we obtain depend on the value of the
viscosity parameter $\alpha$. For standard
values, $0.1\la \alpha \la 0.3$, so much mass evaporates from the disk
into the hot corona that the additional mass flow from the outside is
a negligible contribution. The evaporation then with time could lead to
the complete disappearance of the original disk. Only for values $\alpha
\la 0.07$ there might be solutions in which the incoming accreted gas
condenses into the disk. However, the vertical structure computations
show that in all these solutions the calculated
Bremsstrahlung of the corona in the X-ray band by far exceeds the
observed luminosity.

From this we conclude that, if our modeling is correct and applicable,
at present no cool disk around the Galactic Center exists. We
shortly discussed what limit this puts on an inert disk that might have
originally remained from a phase of star formation in which the
young bright stars presently seen close to the Galactic Center were
formed. We also discuss the difference to the alternative non-radiative
condensation model of Nayakshin (2004).

\begin{acknowledgements}
B.F.Liu would like to thank the Alexander von Humboldt-Foundation for support.
\end{acknowledgements}

\end{document}